\documentclass[twocolumn]{autart}

\usepackage{mathrsfs}
\usepackage{amsmath,amssymb,amsfonts}
\usepackage{algorithmic}
\usepackage{graphicx}
\usepackage{textcomp}
\usepackage{stfloats}
\usepackage{subfigure}
\usepackage{natbib}
\usepackage[pdftex,bookmarks=true,bookmarksnumbered=true,bookmarksopen=true,
colorlinks=true,citecolor=blue,linkcolor=red,anchorcolor=green,urlcolor=blue]{hyperref}

\newtheorem{coro}{Corollary}
\newtheorem{rema}{Remark}
\newtheorem{lemma}{Lemma}

\newcommand{\real}{\mathcal{R}}

\begin{document}

\begin{frontmatter}
\title{Leaderless Consensus of a Hierarchical Cyber-Physical System}

\thanks[footnoteinfo]{This paper was not presented at any IFAC 
meeting.
}

\author[zju]{Xiao Chen}\ead{chenxiao\_7@zju.edu.cn},
\author[zucc]{Yanjun Li}\ead{liyanjun@zucc.edu.cn},
\author[zju]{Arman Goudarzi}\ead{agoudarzi@zju.edu.cn},
\author[zju]{Ji Xiang}\ead{jxiang@zju.edu.cn}

\address[zju]{College of Electrical Engineering, Zhejiang University, PR China}
\address[zucc]{Zhejiang University City College, PR China}


\begin{abstract}
This paper models a class of hierarchical cyber-physical systems and studies its associated consensus problem. The model has a pyramid structure, which reflects many realistic natural or human systems. By analyzing the spectrum of the coupling matrix, it is shown that all nodes in the physical layer can reach a consensus based on the proposed distributed protocols without interlayer delays. Then, the result is extended to the case with interlayer delays. A necessary and sufficient condition for consensus-seeking is derived from the frequency domain perspective, which describes a permissible range of the delay. Finally, the application of the proposed model in the power-sharing problem is simulated to demonstrate the effectiveness and significance of the analytic results.
\end{abstract}

\end{frontmatter}

\section{Introduction}
Advanced technologies in communication and computation techniques have been widely used in the information sensing, control and operation of many physical systems such as power networks, medical devices and manufacturing equipment\citep{Yu2016,Gatouillat2018,Monostori2016}. Such systems that connect the cyber world to the physical world are called cyber-physical systems (CPS), which are characterized by tightly coupling between computation, communications and physical processes \citep{Antsaklis2014}.

Among CPSs, there is a type of system that integrates a finite number of subsystems in a hierarchical structure. The subsystems collectively work together to achieve desired global goals. For instance, consider a power network with many transmission system operators (TSOs). A TSO manages the local generators to supply customers without depending much on the neighbouring TSOs; meanwhile, the TSOs interconnect a broader region by sharing aggregate information for reliability and economic reasons \citep{Wen2017}. Another example is the hierarchical multi-agent system, where all agents are divided into several groups in the first layer \citep{Nguyen2015}. The agents perform local actions in the lower layer and exchange information to attain a cooperative purpose in the upper layer.

\begin{figure}
	\centering
	\includegraphics[width=1\linewidth]{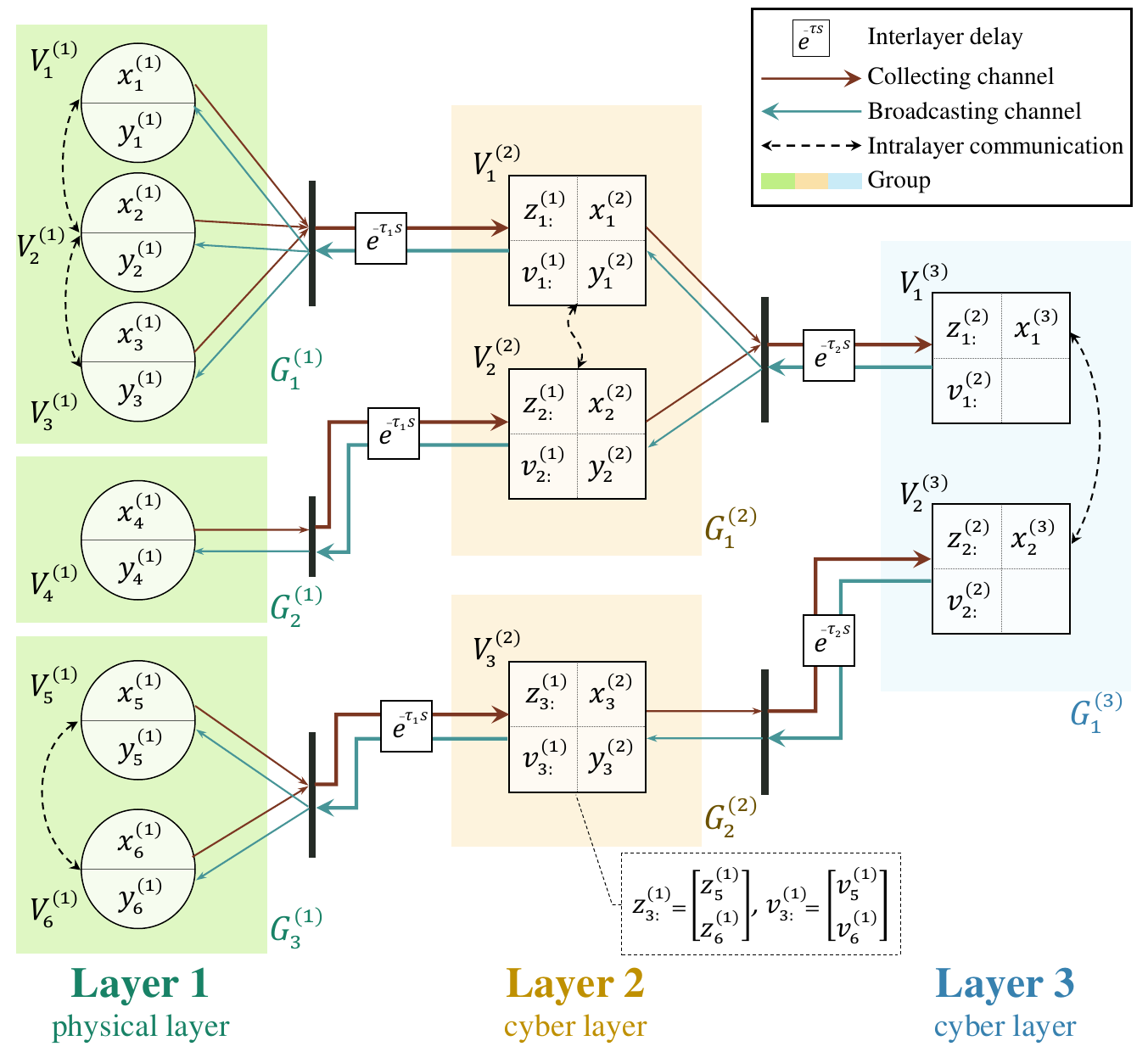}
	\caption{Pyramid hierarchical structure.}
	\label{hierarchy}
\end{figure}

The above-mentioned examples can be sketched into a hierarchical pyramid structure as shown in Fig.\ref{hierarchy}, where the physical system is abstracted as the nodes at the bottom. These physical nodes are coupled into different groups and it is possible to further combine them into larger clusters through coupling them with upper cyber nodes. Finally, all physical nodes are indirectly coupled through the hierarchical structure.
The nodes of higher layers have aggregate information about their subordinate groups. This is completely different from multiagent systems with peer-to-peer information exchange. How to model such hierarchical pyramid structures to achieve certain coordination is a challenging problem.

The consensus problem as a fundamental challenge in distributed control and coordination, has attracted extensive attention in the last decade and encouraged a number of researchers to work on the consensus-based applications \citep{Olfati2004,Olfati2007,Nedich2015,Yang2016,Xiang2017}.
There are also several works with the consensus of the hierarchical network.
In the field of leader-following consensus, it is shown that the hierarchical network can achieve a fast convergence rate of consensus \citep{Shao2016}, which aligns with the phenomenon of pigeon flock or swarm intelligence \citep{Nagy2010}.
Several studies have been conducted on the application of hierarchical structure in the area of leaderless consensus-based problems. Smith et al. introduced a hierarchical cyclic pursuit scheme where all agents are placed in the cyclic pursuit within each group. At the same time, the centroid of each group is following the centroid of the next group in a sequential manner \citep{Smith2005}.
Most of the previous studies only considered homogeneous gains, however, \cite{Mukherjee2016} presented a new method by taking the heterogeneous gains into account and generalized its convergence properties. Since the aforementioned hierarchical cyclic pursuit scheme failed to describe the weakness of intergroup couplings in the real world, \cite{Tsubakino2012} proposed the concept of low-rank interactions. In \cite{Iqbal2018}, a Cartesian product based hierarchical scheme is proposed, which does not necessarily exhibit circulant symmetry as required in the hierarchical cyclic pursuit method. Based on the Lyapunov function method, several researchers proposed sufficient conditions for the consensus of a hierarchical multi-agent system with interlayer communication delay \citep{Duan2015}. However, the discussed previous studies on leaderless consensus cannot be applied in the pyramid structure, and moreover, they are restricted to the two conditions: either the communication graphs of the subgroups which are located in the same layer must be identical, or a special circulant matrix is required.

This paper formulates a general mathematical model for a hierarchical pyramid CPS to break through the two above-mentioned restricted conditions and investigates its related consensus problems.
The first layer of the proposed model is the physical layer, where physical systems are restricted as first-order integrators.
The other layers are hierarchical cyber layers for computation and communication.
Interlayer communication delays are considered in the proposed model.
%One of the applied scenarios of the proposed method is to solve the power-sharing problem in the power system. The generators can be regarded as the nodes in the physical layer, and the dispatch organizations (DOs) at multiple levels can be regarded as the nodes in the hierarchical pyramid cyber layers. To be specific, the DOs in the second layer is responsible for inter-provincial power dispatch, while the DOs in the third layer perform power dispatch in a larger area such as different regions.

The major contributions of this study can be listed as follows:
\begin{itemize}
	\item [1)] Presenting a hierarchical model with distributed consensus protocols. It is closer to the pyramid structure of human society, and its subgroups are allowed to have different communication graphs.
	\item [2)] Providing a necessary and sufficient condition for the consensus of the hierarchical CPS with interlayer delays.
	\item [3)] Applying the proposed model to solve the power-sharing problem in the power system.
\end{itemize}

The remainder of this paper is organized as follows: Section \ref{chap02} constructs the mathematical formulation of the proposed hierarchical CPS. Section \ref{chap03} presents distributed protocols. 
Section \ref{chap04} analyzes the convergence properties of the hierarchy model without the interlayer delay, while section \ref{chap05} takes this interlayer delay into the account and presents a necessary and sufficient condition for the consensus. 
The simulation results of the proposed model are given in section \ref{chap06}.
Finally, section \ref{chap07} concludes the paper.
All the proofs are placed in Appendix \ref{appen_theorem}.

\section{Mathematical formulation}\label{chap02}

\subsection{Background in Laplacian matrix}
Consider an undirected graph $G\!=\!\{V,E\}$, where $V$ is the set of nodes and $E$ is the set of edges denoted by $(i,j)$. Two nodes $u$ and $v$ of $G$ are neighbors if $\{u,v\}\in E$. The binary adjacency matrix of $G$ is the non-negative matrix $A$, where $a_{ij}=1$ if $(i,j)\in E$, and $a_{ij}=0$ otherwise. The out-degree matrix $D_{out}$ is a diagonal matrix defined as ${D_{out}=diag(A\textbf{1})}$, where $\textbf{1}$ is the column vector with compatible dimensions and all components being 1. Laplacian matrix of $G$ is given by ${L=D_{out}-A}$, which has the following properties: ($i$) its row-sums are zero, ($ii$) its diagonal entries are non-negative, and ($iii$) its non-diagonal entries are non-positive \citep{Bullo2017}.

\subsection{Hierarchical structure}
We first review an example of a hierarchical power system in conjunction with Fig.\ref{hierarchy}. The generators can be regarded as the nodes in the physical layer, and the dispatch organizations (DOs) at multiple levels can be regarded as the nodes in the hierarchical pyramid cyber layers. To be specific, the DOs in the second layer are responsible for inter-city power dispatch, while the DOs in the third layer perform power dispatch in a larger area such as different provinces.

Motivated by the above example, we assume that the proposed hierarchical CPS with pyramid structures satisfy the following rules:
\begin{enumerate}
	\item [r1)] The graph of each group should be connected and there is no communication link between the groups of the same layer.
	\item [r2)] The physical layer is at the bottom, which has the largest number of nodes, while the top layer has only one group with the smallest number of nodes. 		
	\item [r3)] Each group has one superior node in the next layer, except for the group in the top layer; each node has a subordinate group in the last layer, except for physical layer nodes. 
\end{enumerate}
The first rule and second rule shows that the isolated groups are ultimately connected through the top layer nodes. The third rule implies the number of nodes decreases as the number of layer increases, thereby forming a pyramid structure. The extreme case is that all the groups, except for the top layer group, have only one node so that the number of nodes in all layers is the same.

\subsection{Dynamical models}\label{parameters}
Consider a proposed hierarchical CPS with $M$ layers,
where the $l$-$th$ layer contains $N^{(l)}$ nodes, $l=1,2,\cdots,M$. The first layer is the physical layer and the other layers are the cyber layers. The physical layer consists of $N^{(1)}$ dynamic nodes, modeled by
\begin{equation}\label{physical_dynamic}
\dot{x}^{(1)}_i = u_i, \quad i=1,2,\cdots, N^{(1)},
\end{equation}
where $x^{(1)}_i \in \real$ is the state value and $u_i$ is the input. The $N^{(l)}$ nodes in the $l$-th layer form $N^{(l\!+\!1)}$ groups, and each group is the subordinate group of one node in the $(l+1)$-th layer. Consequently, all nodes in the top layer form only one group, namely $N^{(M+1)}=1$.

The $p$-th group in the $l$-th layer contains $k_p^{(l)}$ nodes, which has an undirected communication graph denoted by $G^{(l)}_{p}$.
All nodes in the same layer are numbered according to the ascending order of the respective group index. Let $V^{(l)}_i$ denote the $i$-th node in the $l$-th layer. The set of neighbors of $V^{(l)}_{i}$ in the $l$-$th$ layer is denoted by $\mathcal{N}^{(l)}_{i}$.

Each group $G_p^{(l)}$ in the $l$-th layer has the unique superior node $V^{(l+1)}_p$, and $G_p^{(l)}$ is called the subordinate group of node $V^{(l+1)}_p$. Use $V^{(l+1)}_p=\bar{G}_p^{(l)}$ to represent that $V^{(l+1)}_{p}$ is the superior node of $G^{(l)}_{p}$. The superior node $V_p^{(l+1)}$ collects and broadcasts the information from and to its subordinate group $G_p^{(l)}$. This action represents the interlayer communication link, which is located between the superior node and the node of its subordinate group.

Similar to the definition of receptive field in convolutional neural networks, the receptive field of $V^{(l)}_i$ in the hierarchical structure refers to the physical nodes that are visible to $V^{(l)}_i$. The physical number of $V^{(l)}_i$, denoted by $n_{i}^{(l)}$, is the number of physical nodes in its receptive field. If each physical node has a nonnegative physical weight, we define the physical weight $a_{i}^{(l)}$ of the cyber node $V^{(l)}_i$ as the sum of the physical weights of the physical nodes in its receptive field.

Generally, it is reasonable to assume that nodes in the same layer communicate immediately, whereas the nodes between different layers communicate with delay. Assume that all interlayer delays arising from the $l$-th layer to the $(l+1)$-th layer are the same, which is denoted by $\tau_{\,l}$.
In this sense, we can formulate the information exchanged between $V^{(l)}_{i}$ and its superior node,
\begin{multline}\label{cyber_dynamic}
\left\{
\begin{split}
z_{i}^{(l)}(t)=x^{(l)}_{i}(t-\tau_{\,l})\\
y_{i}^{(l)}(t)=v^{(l)}_{i}(t-\tau_{\,l})
\end{split}
\right., \quad l=1,\cdots,M-1,\\
i=1,\cdots, N^{(l)},
\end{multline}
where $x_{i}^{(l)}$ is the information sent from $V^{(l)}_{i}$, which is collected by its superior node $V^{(l+1)}_p$ as $z_{i}^{(l)}$. The received message $y_{i}^{(l)}$ of $V^{(l)}_{i}$ is broadcasted from its superior node as $v_{i}^{(l)}$.
The information collected by $V^{(l+1)}_p$ from its subordinate group $G^{(l)}_p$ can be written in a column vector $z_{p:}^{(l)}=[z_{i}^{(l)}]^T\in\real^{k_p^{(l)}}$ for $V^{(l)}_i \in G^{(l)}_p$. Similarly, we use $v_{p:}^{(l)}=[v_{i}^{(l)}]^T\in\real^{k_p^{(l)}}$ to denote the information broadcasted by $V^{(l+1)}_p$ to its subordinate group $G^{(l)}_p$.
\begin{rema}
	\emph{Consider the situation where all nodes are clock synchronized. Given the time of sending information and setting an appropriate update interval $\tau$ for all receiving nodes, the interval $\tau$ is greater than the maximum communication delay. In this way, the information used for updating protocols is delayed by equal time $\tau$. Similar ideas can be found in \cite{Zhang2010, Zhang2005}, where queues are added to receiver nodes to handle random delays. From this perspective, it is reasonable to assume an equal delay between neighboring layers.}
\end{rema}

\subsection{Three-layer example}
In this section, we present a three-layer example as shown in Fig.\ref{hierarchy} to illustrate the parameters and contents introduced in section \ref{parameters}.
In the physical layer, there are $V^{(1)}_1$, $V^{(1)}_2$, $V^{(1)}_3$, $V^{(1)}_4$, $V^{(1)}_5$ and $V^{(1)}_6$ from top to bottom. These nodes are divided into three groups, and each group is the subordinate group of one node in the second layer. For instance, the third group in the first layer, represented by $G^{(1)}_3$, is the subordinate group of the third node $V^{(2)}_3$ in the second layer. We call $V^{(2)}_{3}$ the superior node of $G^{(1)}_3$, denoted by $V^{(2)}_{3}=\bar{G}^{(1)}_3$.

An intralayer edge connects $V^{(1)}_{5}$ and $V^{(1)}_{6}$ in $G^{(1)}_3$, but there is no intralayer edge between different groups in the same layer.
The interlayer communication links are located between all nodes in every group and their superior nodes. The receptive field of $V^{(2)}_1$ is group $G_1^{(1)}$ and the physical number satisfies $n^{(2)}_1=3$. All nodes in $G^{(1)}_1$ and $G^{(1)}_2$ are the receptive field of $V^{(3)}_{1}$, so the physical number of $V^{(3)}_{1}$ is $n^{(3)}_{1}=4$.

The node $V^{(2)}_{3}$ has interlayer edges with its subordinate group $G^{(1)}_3$ and its superior node $V^{(3)}_{2}$.
Let $x^{(1)}_{6}$ denote the information sent from $V^{(1)}_{6}$ to $V^{(2)}_{3}$. If we take the interlayer delay into account, then the information received by $V^{(2)}_{3}$ can be expressed by $z^{(1)}_{6}(t)=x^{(1)}_{6}(t-\tau_1)$, where $\tau_1$ is the delay time of interlayer communication between the first layer and the second layer. Similarly, we can obtain $y^{(1)}_{6}(t)=v^{(1)}_{6}(t-\tau_1)$, which describe the interlayer delay when the information is broadcasted from $V^{(2)}_{3}$ to $V^{(1)}_{6}$. Meanwhile, $x^{(2)}_{3}$ and $y^{(2)}_{3}$ are the information of $V^{(2)}_{3}$ which are exchanged with its superior node $V^{(3)}_{2}$.

\subsection{Statement of the Problem}

Based on the presented communication structure, several assumptions are made as follows:
\begin{enumerate}
	\item [A1)] The available information of the physical node $V^{(1)}_i$ includes the neighboring node state $x^{(1)}_j$ with $V^{(1)}_j \in \mathcal{N}^{(l)}_{i}$ and the information $y^{(1)}_{i}$ broadcasted from its superior node.
	\item [A2)] The available information of cyber layer node $V^{(l)}_p$ includes the neighboring node state $x^{(l)}_{r}$ with $V_{r}^{(l)}\in\mathcal{N}^{(l)}_{p}$, the collected information $z_{p:}^{(l-1)}\in\real^{k_p^{(l-1)}}$ from its subordinate group and the broadcasted information $y_p^{(l)}$ from its superior node. Here $y_p^{(M)}$ is null and $l=2,\cdots,M$.
	\item [A3)] Every group has a connected graph.
\end{enumerate}

Concerning the stated assumptions, the main goal of the study can be formulated as designing the following protocols:
\begin{enumerate}
	\item [1)] a control protocol $u_i=u(x^{(1)}_i,x^{(1)}_j,y^{(1)}_i)$ with $V_{j}^{(1)}\in\mathcal{N}^{(1)}_{i}$,
	\item [2)] a collecting protocol $x_{p}^{(l)}=x(z^{(l-1)}_{p:})$,
	\item [3)] and a broadcasting protocol $v_{i}^{(l-1)}=v(x_{p}^{(l)},x_{r}^{(l)},y_p^{(l)})$ with $V_{i}^{(l-1)} \in G_p^{(l-1)}$ and $V_{r}^{(l)}\in\mathcal{N}^{(l)}_{p}$,
\end{enumerate}
to solve the consensus problem in the proposed hierarchical CPS. It is worth noting that this research can provide theoretical support for the consensus-based applications in hierarchical structures, such as the power-sharing problem in hierarchical power systems.

\section{Protocols}\label{chap03}
The proposed control protocol is
\begin{equation}\label{control}
u_i=-\dfrac{1}{a_{i}^{(1)}}\sum_{V_{j}^{(1)}\in\mathcal{N}^{(1)}_{i}}(x^{(1)}_i-x^{(1)}_j)+y_i^{(1)}, \ i=1,\cdots, N^{(1)},
\end{equation}
This protocol helps the nodes to reach a consensus within their group. By the received information $y^{(1)}_i$, the state values for isolated groups in the physical layer can converge to a common value.

For the superior node $V^{(l)}_{p}$ to obtain the weighted average of its subordinate group $G^{(l-1)}_p$, the collecting protocol is given by
\begin{equation}\label{collecting}
x_{p}^{(l)}={C_p^{(l-1)}} z_{p:}^{(l-1)}, \ l=2,\cdots, M, \ p=1,\cdots, N^{(l)},
\end{equation}
where $C_p^{(l-1)}$ is a $k_p^{(l-1)}$-dimensional row vector with non-negative entries that can add up to one.

The broadcasting protocol can be written as
\begin{multline}\label{broadcasting}
v_{i}^{(l-1)}=-\dfrac{1}{a_{p}^{(l)}}\sum_{V_{r}^{(l)}\in\mathcal{N}^{(l)}_{p}}(x_{p}^{(l)}-x_{r}^{(l)})+y_p^{(l)},\\
l=2,\cdots, M,\qquad	p=1,\cdots, N^{(l)},
\end{multline}
where $V_i^{(l-1)}\in G_p^{(l-1)}$.
It is clear that all nodes in ${G}_p^{(l-1)}$ receive the same message from their superior node $V^{(l)}_p$.

\begin{rema}
\emph{By modelling the collection and broadcast of information in human society through two proposed protocols, the proposed model can solve the consensus problem in the more practical pyramid structure. The model allows subgroups with the different communication graphs instead of with the same communication graphs as required in \cite{Mukherjee2016}, \cite{Tsubakino2012} and \cite{Iqbal2018}. Moreover, the kronec product structure in \cite{Tsubakino2012} and the circulant matrix in \cite{Mukherjee2016} are no longer required.}
\end{rema}

Equations (\ref{physical_dynamic})-(\ref{broadcasting}) formulate the proposed hierarchical cyber-physical system, which can be illustrated by using an electrical power network. A DO in the cyber layer collects regional generation capacity and the required load demand information to control the power output of generators that are located in the physical layer. In power system operation, the power-sharing problem aims to force the output of generators to reach the same ratio concerning their maximum output power. Notation $x^{(1)}_i$ represents the output ratio of a generator, which can be affected by the state of the neighboring node $x^{(1)}_j$ and the received information $y^{(1)}_i$ from its DO. These DOs share the aggregate information $x^{(2)}_p$ with their neighbors and coordinate power transfer between regions by sending $v^{(1)}_{i}$ to regional generators. The issue of power balance in this model will be discussed in section \ref{chap06}.

Let $x^{(l)}$, $y^{(l)}$, $z^{(l)}$ and $v^{(l)}$ be column vectors with entries $x^{(l)}_i$, $y^{(l)}_i$, $z^{(l)}_i$ and $v^{(l)}_i$, respectively ($i=1,\cdots,N^{(l)}$). Then, the proposed hierarchical system can be expressed in a compact form. The node dynamics in the physical layer is given by
\begin{equation}\label{hie_formu1_vec}
\dot {x}^{(1)}=-K^{(1)}L^{(1)}_{D}x^{(1)}+y^{(1)},
\end{equation}
and the process in the cyber layer can be formulated by
\begin{equation}\label{hie_formu2_vec}
\left\{
\begin{split}
z^{(l-1)}(t)&=x^{(l-1)}(t-\tau_{l\!-\!1})\\
x^{(l)}(t)&=C^{(l-1)}\cdot z^{(l-1)}(t)\\
v^{(l-1)}(t)&=B^{(l-1)}\cdot [-K^{(l)}L^{(l)}_{D}\cdot x^{(l)}(t)+y^{(l)}(t)]\\
y^{(l-1)}(t)&=v^{(l-1)}(t-\tau_{l\!-\!1})
\end{split}
\right.,
\end{equation}
where $L^{(l)}_{D}$ is the Laplacian matrix of the $l$-th layer, which can be written by
\begin{equation}
L^{(l)}_{D}=\mathrm{diag}(L^{(l)}_{1},\cdots,L^{(l)}_{N^{(l\!+\!1)}}), \quad l=1,\cdots,M,
\end{equation}
where $L^{(l)}_{p}\in \real^{k^{(l)}_{p}\times k^{(l)}_{p}}$ denote the Laplacian matrix of $G^{(l)}_{p}$. The matrix $K^{(l)}$ is a diagonal matrix represented by
\begin{equation}
K^{(l)}=\mathrm{diag}(a^{(l)}_{1},\cdots,a^{(l)}_{N^{(l)}})^{-1}, \quad l=1,\cdots,M.
\end{equation}
The matrix $B^{(l)}$ is a block diagonal matrix with $N^{(l+1)}$ blocks, describing the information broadcasted from the $(l+1)$-th layer to the $l$-th layer, and given by
\begin{equation}
B^{(l)}=\mathrm{diag}(\textbf{1}_{1:}^{(l)},\cdots,\textbf{1}_{N^{(l+1)}:}^{(l)}), \ l=1,\cdots,M-1,
\end{equation}
where $\textbf{1}_{p:}^{(l)}$ is a $k^{(l)}_{p}$-dimensional column vector with all entries being 1. The matrix $C^{(l)}$ represents the information collection from the $l$-th layer to the $(l+1)$-th layer, which can be written by
\begin{equation}
C^{(l)}=\mathrm{diag}(C_1^{(l)},\cdots,C_{N^{(l+1)}}^{(l)}), \ l=1,\cdots,M-1.
\end{equation}

If the interlayer delay is ignored ($\tau_{\,l} =0$), then hierarchical system which has been defined by (\ref{hie_formu1_vec})-(\ref{hie_formu2_vec}) can be simplified as
\begin{equation}\label{hie_formu_sim}
\dot{x}^{(1)}=-\sum_{l=1}^{M}{L^{(l)}x^{(1)}}=-Lx^{(1)},
\end{equation}
where
\begin{equation}\label{L_definition}
L^{(l)}=\left\{
\begin{split}
&\qquad \quad \quad \ \, K^{(l)}L^{(l)}_{D} \qquad \quad \quad \ \, , \ l=1,\\
&\ \prod_{i=1}^{l-1}B^{(i)}\cdot K^{(l)} L^{(l)}_{D}\cdot \prod_{i=l\!-\!1}^{1}C^{(i)},\ l=2,\cdots,M,\\
\end{split}
\right.
\end{equation}
and
\begin{equation}\label{L_sum}
L=\sum_{l=1}^{M}{L^{(l)}}.
\end{equation}
Here the notation $\prod_{i=l\!-\!1}^{1}\!\!C^{(i)}$ is specified as $C^{(l-1)}\!\times\!  \cdots \!\times\! C^{(1)}$, since matrices product highly depends on the sequence.

\begin{rema}
\em The proposed model is significantly different from the consensus algorithm based on two main reasons. First, instead of the traditional consensus protocol, specific collecting and broadcasting protocols are designed for cyber nodes to make full use of aggregate information in the hierarchical structure. According to the proposed model (\ref{hie_formu2_vec}), there is no dynamic in the cyber nodes. Second, as $L$ given by (\ref{L_sum}) does not portray a Laplacian matrix, many well-known results cannot be directly used to analyze the proposed model.
\end{rema}

\section{Hierarchy model with no interlayer delay}\label{chap04}

In this section, we analyze the related consensus problems of the proposed hierarchical system (\ref{hie_formu_sim}) without delay. Throughout this paper, notation $\Lambda(X) $ denotes the eigenvalues set of $X$, and $\textbf{1}_{x}$ denotes the column vector with $x$-dimensions and all components being $1$.

Before discussing the obtained results of this study, we introduce a preliminary Lemma.

\begin{lemma}\label{lemma02}
	Given two diagonal matrices $B=\mathrm{diag}(\textbf{1}_{m_1},$ $\cdots,\textbf{1}_{m_n})$ and $C=\mathrm{diag}(C_1,\cdots,C_n)$ satisfying $CB=I_{n\times n}$, where $C_i$ is a $m_i$-dimensional row vector and $\sum_{i=1}^n m_i = m$. Then for any matrix $A\in\real^{n\times n}$ with $A\textbf{1}=\textbf{0}$, and any diagonal matrix $E=\mathrm{diag}(E_1, \cdots, E_n)$ with $E_i\in\real^{m_i\times m_i}$ and $E_i\textbf{1}_{m_i}=\textbf{0}$, the eigenvalues set of matrix ${F}=E+BAC$ satisfies $[\Lambda(E)\cup \Lambda(A)]\subset\Lambda(F)$.
\end{lemma}
Let $L^{(l)}_{e}$ be a scaled matrix of $L^{(l)}_{D}$, given as
\begin{equation}\label{Lapequa}
L^{(l)}_{e}=K^{(l)}{L^{(l)}_{D}},
\end{equation}
then we can construct the following matrix sequence based on equation (\ref{L_definition}),
\begin{equation}\label{L_sequence}
\begin{split}
	L^{(M)}_0&=L^{(M)}_e\\
	L^{(M)}_k&=L^{(M\!-\!k)}_e+B^{(M\!-\!k)}L^{(M)}_{k\!-\!1}C^{(M\!-\!k)},\ k=1,\!\cdots\!,M\!-\!1
\end{split}.
\end{equation}

\begin{thm} \label{th02}
	Assume that in the top layer $M$, $N^{(M)}$ nodes form one connected graph, namely $N^{(M+1)}=1$, then for $k=1,2,\cdots,M-1$,
	\begin{equation}		
	\Lambda(L^{(M)}_k)=0\ \cup\ \Lambda(L^{(M-k)}_e)/0\ \cup\ \Lambda(L^{(M)}_{k-1})/0,
	\end{equation}
	where $\Lambda(L^{(M)}_{k-1})/0$ denotes the non-zero eigenvalues set of $L^{(M)}_{k-1}$. Furthermore, 0 is a simple eigenvalue of $L^{(M)}_k$, that is, the algebraic multiplicity of 0 is 1.
\end{thm}

Below is an immediate corollary based on Theorem \ref{th02}.
\begin{coro}\label{corol_Ls}
	Note that the matrix $L$ defined in equation (\ref{L_sum}) can also be represented by $L=L^{(M)}_{M-1}$, so
	\begin{equation}\label{L_eig}
		\Lambda(L)=0\ \cup\ \Lambda(L^{(1)}_e)/0\ \cup\ \cdots\ \cup\ \Lambda(L^{(M)}_e)/0.
	\end{equation}
	To be further, let $L_s=\sum_{l=1}^M s_l L^{(l)}$, where $s_l$ is a complex number except for zero, then
	\begin{equation}
		\Lambda(L_s)=0\ \cup\ \Lambda(s_1L^{(1)}_e)/0\ \cup\ \cdots\ \cup\ \Lambda(s_M L^{(M)}_e)/0.
	\end{equation}
\end{coro}

Corollary \ref{corol_Ls} implies that we can analyze $\Lambda(L^{(l)}_e)$ instead of studying $\Lambda(L)$ directly. Note that ${K^{(l)}}^{-\frac{1}{2}}L^{(l)}_{e}{K^{(l)}}^{\frac{1}{2}}={K^{(l)}}^{\frac{1}{2}}{L^{(l)}_{D}{K^{(l)}}^{\frac{1}{2}}}=Z$ is a real symmetric matrix. One has that $\Lambda(L^{(l)}_{e})=\Lambda(Z)\subset\real$ . 
Also notice that $L^{(l)}_{e}$ is a weighted Laplacian matrix, thus, the real part of eigenvalues of $L^{(l)}_{e}$ is not less than zero. Furthermore, the real part of all other eigenvalues of $L$ is positive, except for the single zero eigenvalue.

%Below is an immediate corollary based on Theorem \ref{th02} and its proof.
%\begin{coro}\label{corol_Ls}
%	Let $L_s=\sum_{l=1}^M s_l L^{(l)}$, where $s_l$ is a complex number except for zero, then
%	\begin{equation}
%	\Lambda(L_s)=0\ \cup\ \Lambda(s_1L^{(1)})/0\ \cup\ \cdots\ \cup\ \Lambda(s_M L^{(M)})/0.
%	\end{equation}
%\end{coro}

\begin{thm}\label{prop1}
	For the hierarchical system (\ref{hie_formu_sim}) without interlayer delay, all nodes in the physical layer asymptotically can reach a consensus given by
	\begin{equation}\label{steady_value_static}
	\lim\limits_{t\to \infty}x^{(1)}(t)=\dfrac{\textbf{1}\textbf{1}^T K}{\textbf{1}^T\cdot K\cdot \textbf{1}}x^{(1)}_0,
	\end{equation}
	where $x^{(1)}_0$ is the initial value of $x^{(1)}$ and
	\begin{equation}\label{K}
	K=diag(a^{(1)}_1,\cdots,a^{(1)}_{N^{(1)}}).
	\end{equation}
	In particular, all physical layer nodes will reach an average consensus if $a^{(1)}_i=a^{(1)}_j$ for $i,j\in\{1,\cdots,N^{(1)}\}$.
\end{thm}

\section{The influence of interlayer delay}\label{chap05}
The main goal of this section is to analyze the impact of time constants on the convergence properties of the proposed system. First, the three-layer example from the perspective of the frequency domain is studied, then the results are extended for the generalization of the presented model.

\subsection{Three-layer example}

Based on the proposed protocols in section \ref{chap03}, the three-layer hierarchical system (as depicted in Fig.\ref{hierarchy}) can be described by
\begin{equation}\label{hierarchy_example}
\left\{
\begin{split}
\dot {x}^{(1)}(t)&=-K^{(1)}L^{(1)}_{D}\cdot x^{(1)}(t)+y^{(1)}(t)\\
x^{(2)}(t)&=C^{(1)}\cdot x^{(1)}(t-\tau_1)\\
y^{(1)}(t)&=B^{(1)}\cdot [-K^{(2)}L_{D}^{(2)}\cdot x^{(2)}(t-\tau_1)+y^{(2)}(t-\tau_1)]\\
x^{(3)}(t)&=C^{(2)}\cdot x^{(2)}(t-\tau_2)\\
y^{(2)}(t)&=B^{(2)}\cdot [-K^{(3)}L_{D}^{(3)}\cdot x^{(3)}(t-\tau_2)]\\
\end{split}
\right..
\end{equation}

If we take the Laplace transform of equation (\ref{hierarchy_example}), then
\begin{equation}\label{freq}
\begin{split}
&sX^{(1)}(s)-x^{(1)}_0\\
=&-L^{(1)}X^{(1)}(s)-\dfrac{L^{(2)}}{e^{2\tau_1 s}}X^{(1)}(s)-\dfrac{L^{(3)}}{e^{2(\tau_1+\tau_2) s}}X^{(1)}(s)
\end{split},
\end{equation}
where $X^{(1)}(s)$ is the Laplace transform of $x^{(1)}$, the initial value of $x^{(1)}$ is noted as $x^{(1)}_0$. Let $L_{\tau,s}=L^{(1)}+\dfrac{L^{(2)}}{e^{2\tau_1 s}}+\dfrac{L^{(3)}}{e^{2(\tau_1+\tau_2) s}}$, then
\begin{equation}\label{freq_solution}
X^{(1)}(s)=[sI+L_{\tau,s}]^{-1} x^{(1)}_0,
\end{equation}
and the characteristic function of equation (\ref{hierarchy_example}) is given by
\begin{equation}\label{charac02}
|sI+L_{\tau,s}|=0.
\end{equation}

\begin{thm}\label{roots}
	For the three-layer hierarchical model (\ref{hierarchy_example}), all nodes in the physical layer asymptotically can reach a consensus given by equation (\ref{steady_value_static}) if and only if all other roots of equation (\ref{charac02}) are in the open left half-plane except for the single root at zero.
\end{thm}

Note that the value of $s$ solved by equation (\ref{charac02}) is the eigenvalue of matrix $-L_{\tau,s}$. Together with the results given in corollary \ref{corol_Ls} we can infer that
\[
\Lambda(L_{\tau,s})=0\ \cup\ \Lambda(L^{(1)}_e)/0\ \cup\  \dfrac{\Lambda(L^{(2)}_e)/0}{e^{2\tau_1 s}}\ \cup\  \dfrac{\Lambda(L^{(3)}_e)/0}{e^{2(\tau_1+\tau_2) s}},
\]
and 0 is a simple eigenvalue of $L_{\tau,s}$. Thus, the solution set of equation (\ref{charac02}) is equal to the union of the solution sets of the following formulas
\begin{subequations}\label{charac02_lamb}
	\begin{align}
	&s=0, \\
	\label{layer1}
	&s+{\lambda}^{(1)}=0, \\
	\label{layer2}
	&s\cdot e^{2\tau_1 s}+{\lambda}^{(2)}=0, \\
	\label{layer3}
	&s\cdot e^{2(\tau_1+\tau_2) s}+{\lambda}^{(3)}=0,
	\end{align}
\end{subequations}
where $\lambda^{(l)}$ denote the non-zero eigenvalue of $L^{(l)}_e$ for $l=\{1,2,3\}$.

Based on the analysis of section \ref{chap04}, we know that $\lambda^{(l)}\in\real$ and the real part of eigenvalues of $L^{(l)}_e$ is not less than zero, so $\lambda^{(l)}\in \real^+$.
\begin{thm}\label{trans}
	For a transcendental equation such as
	\begin{equation}\label{transcendental equation}
	s\cdot e^{\mathrm{T} s}+\lambda=0,
	\end{equation}
	where $s=\sigma+j\omega$ is a complex variable, $\mathrm{T},\lambda\in\real^+$. Then, it has all roots in the open left half-plane if and only if
	\begin{equation}
	\mathrm{T}<\mathrm{T}^\star=\dfrac{\pi}{2\lambda}.
	\end{equation}	
\end{thm}

\begin{rema}
	\emph{A similar conclusion appears in Theorem 10 of \cite{Olfati2004}, however, in this paper, we have provided a new and explicit proof.}
\end{rema}

Below is an immediate Theorem based on Theorem \ref{roots} and \ref{trans}.

\begin{thm}\label{steady_3layer}
	For the three-layer hierarchical model (\ref{hierarchy_example}) with interlayer delay, all physical layer nodes converge to a consensus given by equation (\ref{steady_value_static}) if and only if both of the following formulas hold
	\begin{subequations}\label{stable}
		\begin{align}
		\label{stable_1}
		&\tau_1<\dfrac{\pi}{4\lambda^{(2)}_{max}}, \\
		\label{stable_2}
		&\tau_1+\tau_2<\dfrac{\pi}{4\lambda^{(3)}_{max}},
		\end{align}
	\end{subequations}
	where $\lambda^{(2)}_{max}$ and $\lambda^{(3)}_{max}$ are the maximum eigenvalues of $L^{(2)}_e$ and $L^{(3)}_e$, respectively.
\end{thm}

\subsection{Generalization}
The proposed methods and conclusions in the three-layer example can be extended to establish a general model.

\begin{thm}\label{steady_genera}
	The polynomial characteristics of the hierarchical system (\ref{hie_formu1_vec})-(\ref{hie_formu2_vec}) is equivalent to a series of polynomials:
	\begin{subequations}
		\begin{align}
		&s=0, \\
		&s+\lambda^{(1)}=0, \\
		&s\cdot e^{2s\sum_{i=1}^{l-1}\tau_i} +\lambda^{(l)}=0, \quad l=2,\cdots, M.
		\end{align}
	\end{subequations}
	All nodes in the physical layer asymptotically can reach a consensus given by equation (\ref{steady_value_static}) if and only if
	\begin{equation}
	\sum_{i=1}^{l-1}\tau_i<\dfrac{\pi}{4\lambda^{(l)}_{max}}, \quad l=2,\cdots, M.
	\end{equation}
\end{thm}

\begin{rema}
\emph{Most of the previous studies, such as \cite{Mukherjee2016} and \cite{Iqbal2018}, neglect the communication delays to simplify their analysis. \cite{Duan2015} takes the interlayer delays into account, but it is assumed that communication graphs of the subgroups which are located in the same layer must be identical. Here the effect of delay time has been studied from the perspective of the frequency domain. Corollary \ref{corol_Ls} splits the polynomial characteristic of the hierarchical system into a series of tractable polynomials. In addition, Theorem \ref{trans} shows the influence of interlayer delays on the distribution of characteristic roots.}
\end{rema}

\section{Applications and simulation results}\label{chap06}
The proposed model can be applied to solve the power-sharing problem in the hierarchical power system, in which its purpose is to drive the outputs of the generators to the same ratio with respect to their maximum power output. Taking the three-layer model as shown in Fig.\ref{hierarchy} for example, the six nodes in the physical layer represent six generators, and the nodes in the other two layers represent DOs. The three DOs in the second layer is responsible for inter-city power dispatch, while the two DOs in the third layer perform power dispatch in a larger area such as different provinces. Notation $x^{(1)}_i$ computed by ${p^{(1)}_i}/{\bar{P}^{(1)}_i}$ represents the power ration of a generator, where $p^{(1)}_i$ and $\bar{P}^{(1)}_i$ denotes the output power and maximum output power of the generator, respectively. Let the physical weight of the generator node be equal to its maximum power output, that is $a^{(1)}_i=\bar{P}^{(1)}_i$, then the vector form of the generator output power can be expressed as $p^{(1)}=Kx^{(1)}$, where $K$ is given by equation (\ref{K}). It is assumed that the power output is regulated instantaneously and initially in a state of supply-demand balance, namely $p^{(1)}=p^{(1)}_{ref}$ and $\textbf{1}^T\cdot p^{(1)}(0)=P_D$, where $p^{(1)}_{ref}$ and $P_D$ are the power out reference and the total power demand, respectively. It can be achieved $\textbf{1}^T\cdot \dot{p}^{(1)}(t)=\textbf{1}^T\cdot K\dot{x}^{(1)}(t)=0$ from the proof of Theorem \ref{roots}, thus, the supply-demand balance will not be violated in the transient process. Based on Theorem \ref{steady_3layer}, we can obtain the permissible range for the interlayer delay for the power-sharing in this hierarchical system.

In the following section, the effectiveness of the proposed model will be examined and verified by the simulation results. The aim of case \ref{case01} is to demonstrate that all generators can achieve the power-sharing based on the proposed distributed protocols. 
The other cases (cases \ref{case02}-\ref{case04}) investigate the effect of time delay on the convergence properties of the proposed hierarchical system. Assuming that the maximum power outputs of the generator nodes are $\bar{P}^{(1)}=[0.8,\ 0.7,\ 1.5,\ 1,\ 0.8,\ 1.2]^TMW$. The initial power outputs are $p^{(1)}(0)=[0.24,\ 0.56,\ 0.9,\ 0.9,\ 0.56,\ 0.24]^TMW$ and the total demand is $3.4\ MW$. Therefore, the initial power ratio are $x^{(1)}(0)=[0.3,\ 0.8,\ 0.6,\ 0.9,\ 0.7,\ 0.2]^T$. Let all the edge weights of each graph $G^{(l)}_{p}$ be equal to $1$. Thus, the maximum eigenvalues of $L^{(2)}_e$ and $L^{(3)}_e$ can be listed as $\lambda^{(2)}_{max}=4/3$ and $\lambda^{(3)}_{max}=0.75$, respectively.

\begin{figure}[]
	\centering\subfigure[{\label{sim01-1}}]{		
		\begin{minipage}[b]{0.5\textwidth}			
			\includegraphics[width=1\textwidth]{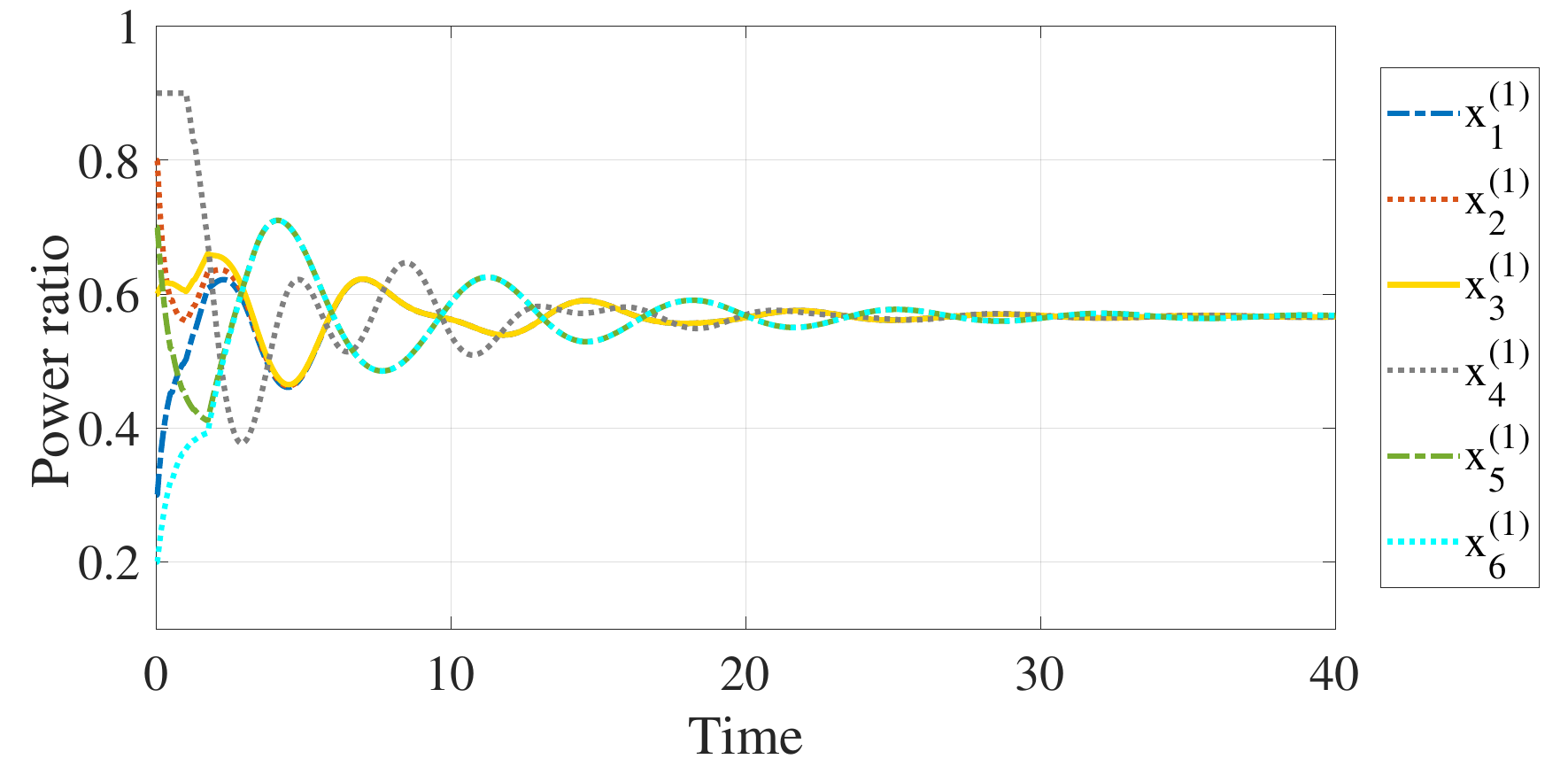}			
		\end{minipage}		
	}
	\subfigure[{\label{sim01-2}}]{		
		\begin{minipage}[b]{0.5\textwidth}			
			\includegraphics[width=1\textwidth]{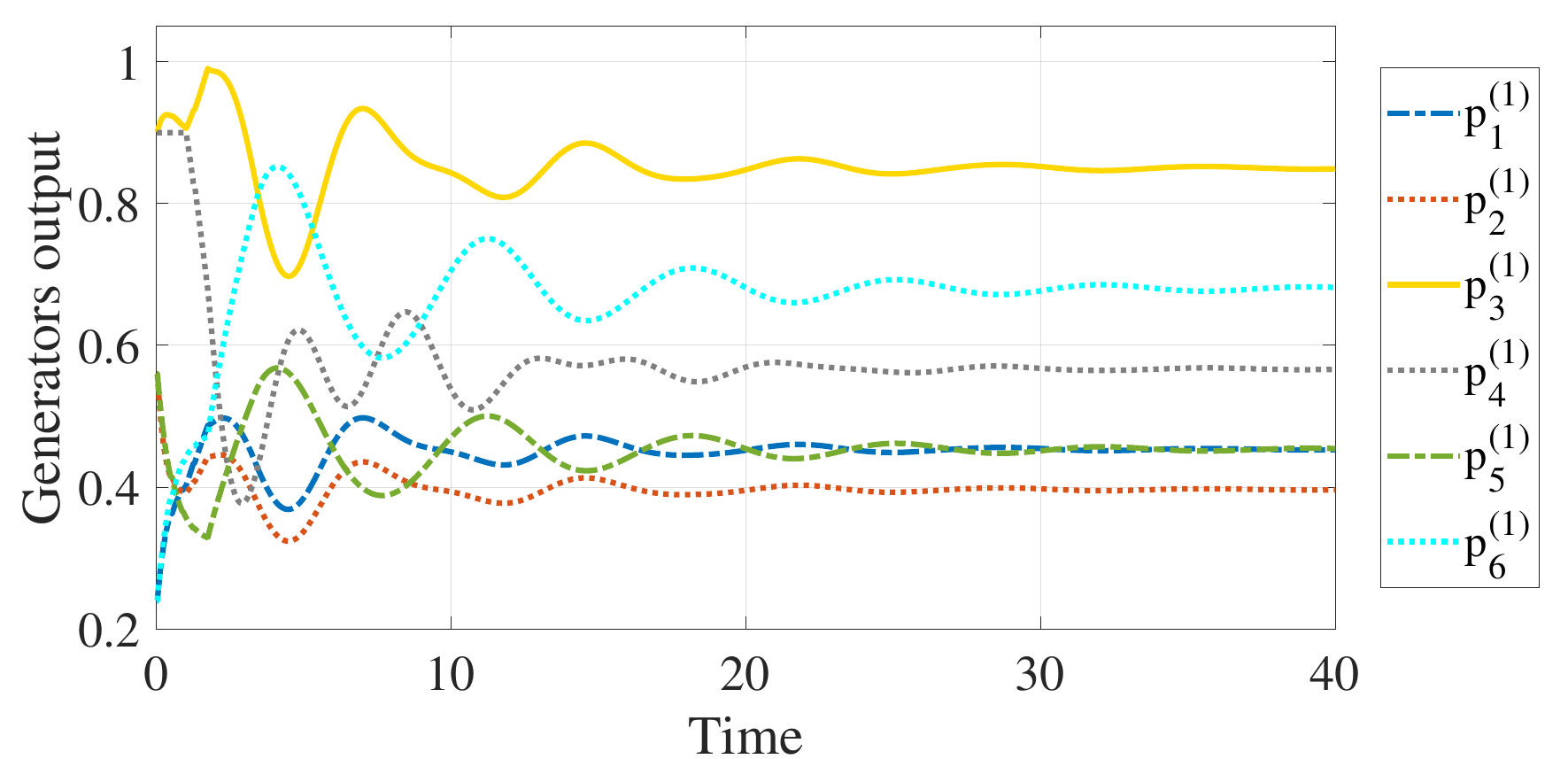}			
		\end{minipage}		
	}
	\subfigure[{\label{sim01-3}}]{		
		\begin{minipage}[b]{0.5\textwidth}			
			\includegraphics[width=1\textwidth]{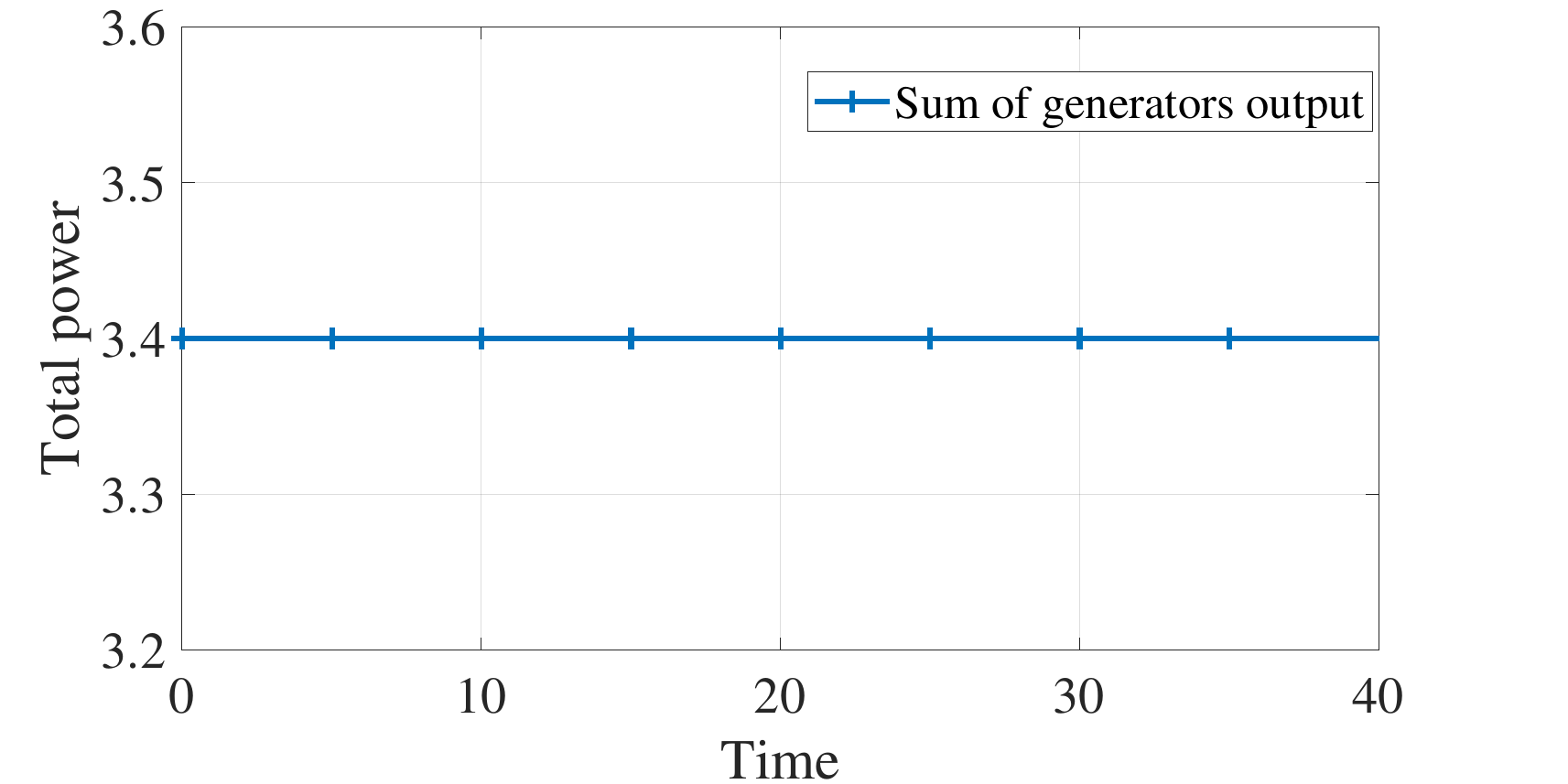}			
		\end{minipage}		
	}
	\caption{power-sharing test when $\tau_1=\frac{\pi}{7}$, $\tau_2=\frac{\pi}{9}$ (Case \ref{case01}).} \label{fig2}	
\end{figure}

\begin{case}\label{case01}
	\emph{$\tau_1=\pi/7$ and $\tau_2=\pi/9$: In this case, equation (\ref{stable}) holds, so the power ratio of all generators converges to a common value $0.5667$ as shown in Fig.\ref{sim01-1}, which is consistent with the result computed by equation (\ref{steady_value_static}). It is clear that $\lim\limits_{t\to \infty}p^{(1)}(t)=K\cdot\lim\limits_{t\to \infty}x^{(1)}(t)=[0.4533,\ 0.3967,\ 0.85,\ 0.5667,\ 0.4533,\ 0.68]^TMW$, which is shown in Fig.\ref{sim01-2}.
	Fig.\ref{sim01-3} indicates that the power balance is maintained from beginning to end.
	It is worth noting that, the $u_i$ will change abruptly at $t=2\tau_1$ and $t=2(\tau_1+\tau_2)$ due to the existence of the interlayer delays, so the non-derivable points appear at the corresponding moment.}
\end{case}

\begin{figure}[]
	\centering	
	\subfigure[$\tau_1=\frac{\pi}{6}$, $\tau_2=\frac{\pi}{6}$.{\label{sim02}}]{		
		\begin{minipage}[b]{0.5\textwidth}			
			\includegraphics[width=1\textwidth]{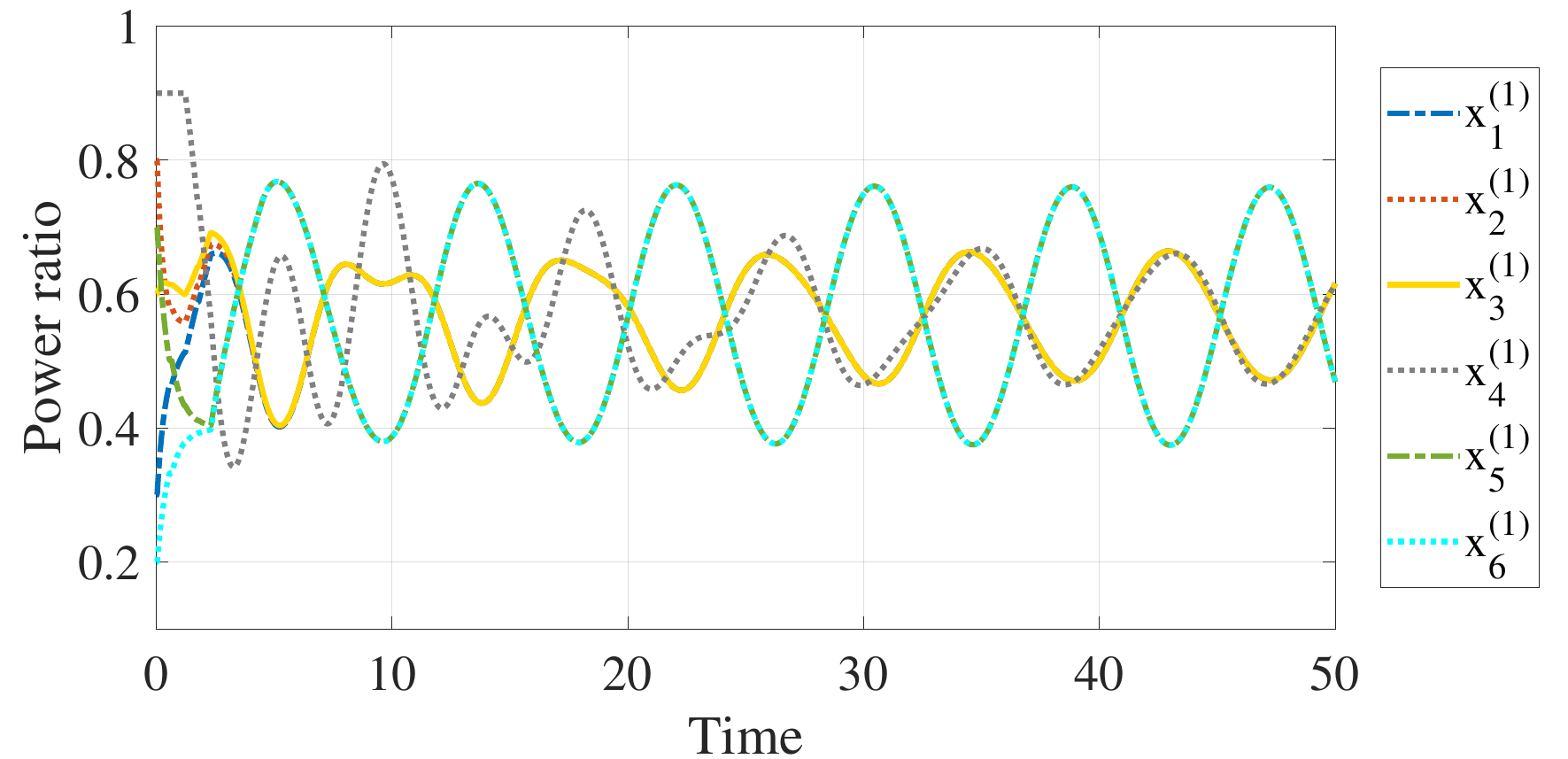}			
		\end{minipage}		
	}
	\subfigure[$\tau_1=\frac{3}{16}\pi$, $\tau_2=\frac{\pi}{12}$.{\label{sim03}}]{		
		\begin{minipage}[b]{0.5\textwidth}			
			\includegraphics[width=1\textwidth]{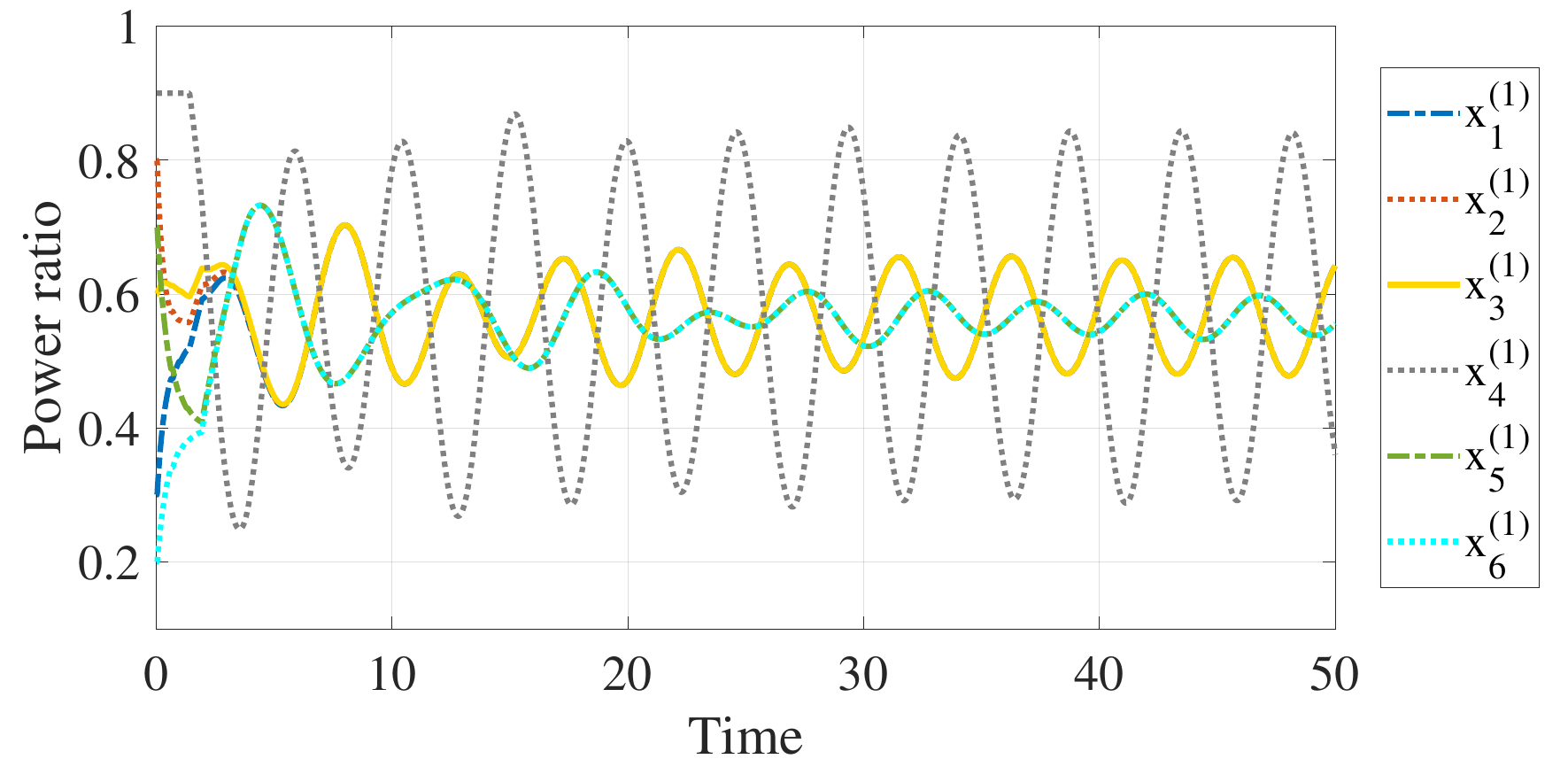}			
		\end{minipage}		
	}
	\subfigure[$\tau_1=\frac{3}{16}\pi$, $\tau_2=\frac{7}{48}\pi$.{\label{sim04}}]{		
		\begin{minipage}[b]{0.5\textwidth}			
			\includegraphics[width=1\textwidth]{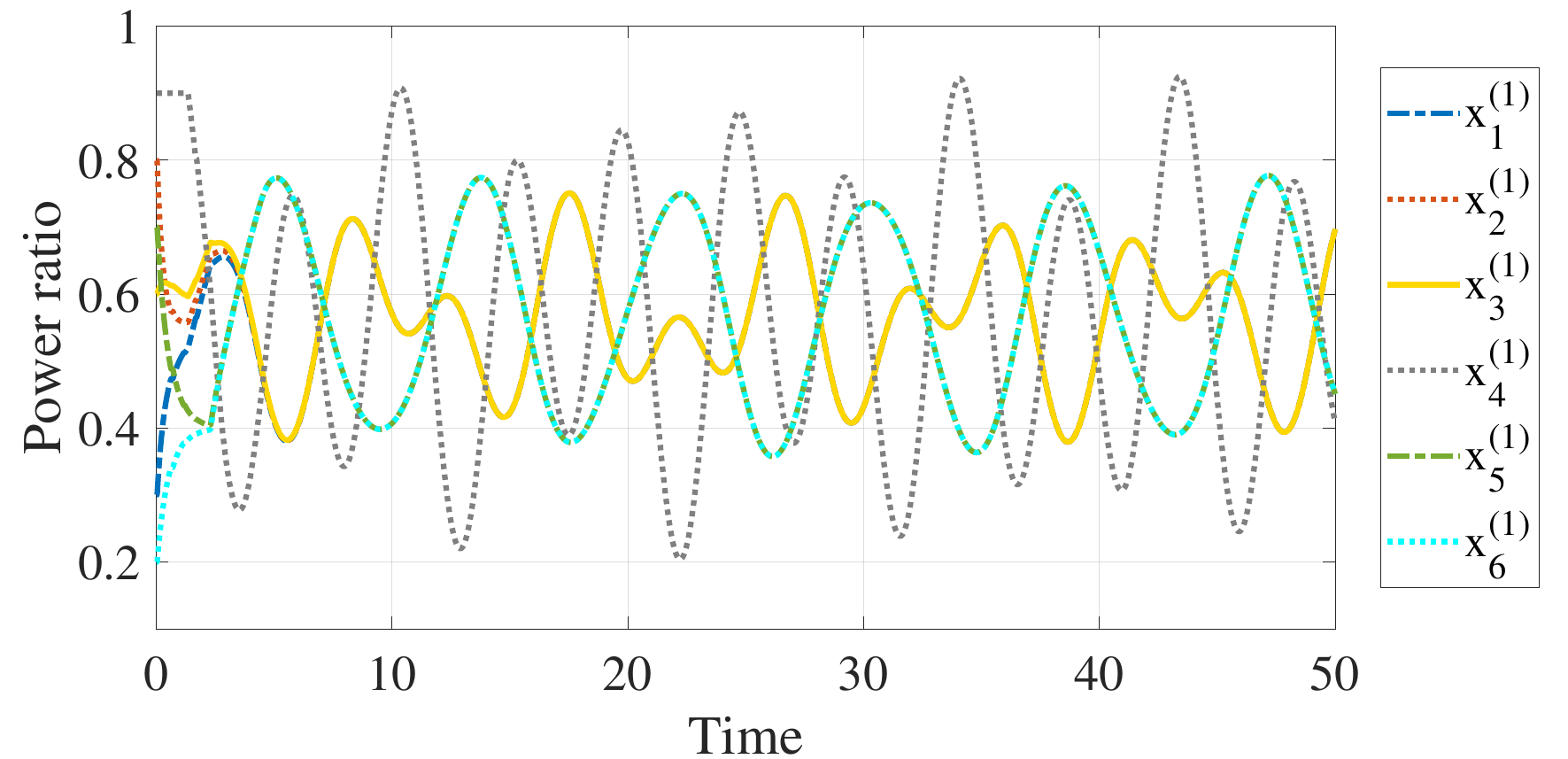}			
		\end{minipage}		
	}
	\caption{Results of the simulation cases (Case \ref{case02} - Case \ref{case04}).} \label{fig3}
\end{figure}

\begin{case}\label{case02}
	\emph{$\tau_1=\pi/6$, $\tau_2=\pi/6$: Fig.\ref{sim02} shows the power ratio trajectories of the generator nodes under this delay time. In this case, equation (\ref{stable_1}) holds but $\tau_1+\tau_2={\pi}/{(4\lambda^{(3)}_{max})}$. It can be inferred from Theorem \ref{trans} that both equations (\ref{layer1}) and (\ref{layer2}) have all roots in the open left half-plane, but equation (\ref{layer3}) has roots on the imaginary axis. Therefore, the system will be in a state of critical oscillation.}
\end{case}

\begin{case}\label{case03}
	\emph{$\tau_1={3}\pi/{16}$, $\tau_2={\pi}/{12}$: In this case, equation (\ref{stable_2}) holds but $\tau_1={\pi}/{(4\lambda^{(2)}_{max})}$.
	So all roots of equations (\ref{layer1}) and (\ref{layer3}) are located in the open left half-plane, but equation (\ref{layer2}) has roots on the imaginary axis. The three-layer example exhibits critical oscillation, as shown in Fig.\ref{sim03}.}
\end{case}

\begin{case}\label{case04}
	\emph{$\tau_1={3}\pi/{16}$, $\tau_2={7}\pi/{48}$: Fig.\ref{sim04} shows the power ratio trajectories of the hierarchical example where $\tau_1={\pi}/{(4\lambda^{(2)}_{max})}$ and $\tau_1+\tau_2={\pi}/{(4\lambda^{(3)}_{max})}$.
	Similarly, we can obtain all roots of equations (\ref{layer1}) that are located in the open left half-plane, but both equations (\ref{layer2}) and (\ref{layer3}) have roots on the imaginary axis. Therefore, at the same time, the system will be in a state of critical oscillation.}
\end{case}

\section{Conclusions}\label{chap07}

In this paper, a hierarchical cyber-physical system has been introduced with distributed consensus protocols that drive the nodes in the physical layer to reach a consensus. For the behavior of the hierarchical model without interlayer delay, we have analyzed its convergence properties. 
%Also, an interlayer communication delay has been addressed with a necessary and sufficient condition that describes a permissible range for the delay time.
Also, a necessary and sufficient condition that describes the delay margin has been provided.
The results of the simulation cases on the power-sharing problem verify the practicality and the effectiveness of the proposed hierarchical model.

\begin{ack}
This work is supported by grants from the Zhejiang Province Natural Science Fund (LY20F030003), the Zhejiang Province Key R\&D Project (2019C01150) and the National Natural Science Foundation of China (61773339).
\end{ack}

\appendix

\section{Proofs of lemmas and theorems}\label{appen_theorem}

%\textbf{\em Proof of the Lemma \ref{lemma01}.}\\[-1.5em]
%
%Let $\lambda$ is any eigenvalue of $A$, that is, $A\textbf{v}_1=\lambda \textbf{v}_1$ for some nonzero vector $\textbf{v}_1$. Let $B\textbf{v}_1=\textbf{v}_2$, then
%	\begin{equation}
%	\begin{split}
%	E \textbf{v}_2&=BAC \textbf{v}_2=B A CB\textbf{v}_1=BA\textbf{v}_1=\lambda B \textbf{v}_1\\
%	&=\lambda \textbf{v}_2
%	\end{split}.
%	\end{equation}
%	Therefore, $\Lambda(A) \subset \Lambda(E)$.
%
%\bigskip

\textbf{\em Proof of the Lemma \ref{lemma02}.}\\[-1.5em]

Let $A\textbf{v}=\lambda_1 \textbf{v}$, due to $CB=I$, then $BACB\textbf{v}=\lambda_1 B\textbf{v}$. Since $E_i \textbf{1}_{m_i}=\textbf{0}$, $E B=\mathrm{diag}(E_1,\cdots,E_n) \mathrm{diag}(\textbf{1}_{m_1},$ $\cdots,\textbf{1}_{m_n})=0$ and therefore $EB \textbf{v}=\textbf{0}$. Thus, 
\[
F\cdot B\textbf{v}=E\cdot B\textbf{v} + BAC\cdot B\textbf{v} = \lambda_1\cdot B \textbf{v},
\]
that is $\Lambda(A) \subset\Lambda(F)$.\\[-1.5em]

It is evident that $0$ is an eigenvalue of $E$ and $F$, noting that $C\textbf{1}_m = \textbf{1}_n$. The remainder is to study the relationship of non-zero eigenvalues between $E$ and $F$. \\[-1.5em]

Let $\textbf{w}$ be the left eigenvector of matrix $E$, that is , $\textbf{w}{'}E = \lambda_2 \textbf{w}{'}$, for some eigenvalue $\lambda_2$. Since $EB=0$, one has $\textbf{w}{'}B=0$. Thus, 
\[
\textbf{w}'F=\textbf{w}'E + \textbf{w}'BAC = \lambda_2 \textbf{w}{'}
\]
that is $\Lambda(E) \subset\Lambda(F)$. To sum up, $[\Lambda(E)\cup \Lambda(A)]\subset\Lambda(F)$.

\bigskip
\textbf{\em Proof of the Theorem \ref{th02}.}\label{proth02}\\[-1.5em]

According to Lemma \ref{lemma02}, it follows that $[\Lambda(L^{(M-k)}_e)\cup \Lambda(L^{(M)}_{k-1})]\subset\Lambda(L^{(M)}_{k})$. This, together with equation (\ref{L_sequence}), yields that any eigenvalue of $L^{(l)}_e$ is the eigenvalue of $L^{(M)}_k$ for $l = \{M-k, M-k+1, \cdots,M\}$. \\[-1.5em]

Since the number of non-zero eigenvalues of $L^{(l)}_e\in \real^{N^{(l)}\times N^{(l)}}$ is $N^{(l)}-N^{(l+1)}$, the number of non-zero eigenvalues of $L^{(M)}_k\in \real^{N^{(M-k)}\times N^{(M-k)}}$ is not less than 
\[
\sum_{l=M-k}^{M}(N^{(l)}-N^{(l+1)})=N^{(M-k)}-N^{(M+1)}=N^{(M-k)}-1.
\] 
This implies that $0$ is a simple eigenvalue of $L^{(M)}_k$ and the algebraic multiplicity of 0 is 1.

\bigskip
\textbf{\em Proof of the Theorem \ref{prop1}.}\\[-1.5em]

This Theorem is a special case of Theorem \ref{roots} where $L_{\tau,s}=L$. For more details, please refer to the proof of Theorem \ref{roots}.

\bigskip
\textbf{\em Proof of the Theorem \ref{roots}.}\\[-1.5em]

Let $e(t)=x^{(1)}(t)-\dfrac{\textbf{1}\textbf{1}^T K}{\textbf{1}^T K \textbf{1}} x^{(1)}_0$, and our goal is to prove $\lim\limits_{t\to \infty}e(t)=0$.
We use $E(s)$ to denote the Laplace transform of $e(t)$, then
\begin{equation}
E(s)=X^{(1)}(s)-\dfrac{1}{s}\cdot \dfrac{\textbf{1}\textbf{1}^T K}{\textbf{1}^T K \textbf{1}}x^{(1)}_0.
\end{equation}
Substituting this equation into equation (\ref{freq_solution}), we can get
\begin{equation}
[sI+L_{\tau,s}]\cdot[E(s)+\dfrac{1}{s}\cdot \dfrac{\textbf{1}\textbf{1}^T K}{\textbf{1}^T K \textbf{1}}x^{(1)}_0]=x^{(1)}_0,
\end{equation}
Since $L_{\tau,s}\cdot \textbf{1}=0$, then
\begin{equation}\label{error}
[sI+L_{\tau,s}]\cdot E(s)=(I-\dfrac{\textbf{1}\textbf{1}^T K}{\textbf{1}^T K \textbf{1}})x^{(1)}_0=e_0,
\end{equation}
where $e_0$ is the initial value of $e(t)$.
Since $\textbf{1}^T\cdot KL_{\tau,s}=\textbf{0}$, then $\textbf{1}^{T}K\dot{x}^{(1)}(t)=0$, so $\textbf{1}^{T}Kx^{(1)}(t)=\textbf{1}^{T}Kx^{(1)}_0$, and we can get
\begin{equation}
\begin{split}
\textbf{1}^{T} K\cdot e(t)&=\textbf{1}^{T} K\cdot [x^{(1)}(t)-\dfrac{\textbf{1}\textbf{1}^T K}{\textbf{1}^T K \textbf{1}} x^{(1)}_0]\\
&=\textbf{1}^{T} K x^{(1)}_0-\textbf{1}^{T} K\dfrac{\textbf{1}\textbf{1}^T K}{\textbf{1}^T K \textbf{1}} x^{(1)}_0=0.
\end{split}
\end{equation}
Take the Laplace transform of the equation above, then $\textbf{1}^{T} K\cdot E(s)=0$. 
Thus equation (\ref{error}) is equivalent to
\begin{equation}
[sI+L_{\tau,s}+\textbf{1}\textbf{1}^TK]\cdot E(s)=e_0,
\end{equation}
that is,
\begin{equation}\label{E_frequ}
E(s)=[sI+L_{\tau,s}+\textbf{1}\textbf{1}^T K]^{-1} e_0,
\end{equation}
so we can get the characteristic equation
\begin{equation}\label{e_charac}
|sI+L_{\tau,s}+\textbf{1}\textbf{1}^T K|=0
\end{equation}
If all roots of equation (\ref{e_charac}) are in the open left half-plane, then $\lim\limits_{t\to \infty}e(t)=0$ holds for any intial states.

The value of $s$ solved by equation (\ref{e_charac}) is the eigenvalues of $-(L_{\tau,s}+\textbf{1}\textbf{1}^T K)$. Since 
\begin{equation}
\lambda(L_{\tau,s}+\textbf{1}\textbf{1}^T K)=\{\textbf{1}^T K \textbf{1}\}\cup\lambda(L_{\tau,s})\backslash\{0\},
\end{equation}
then $\lim\limits_{t\to \infty}e(t)=0$ if and only if all other roots of $|sI+L_{\tau,s}|=0$ are in the open left half-plane except for one at zero.

\bigskip
\textbf{\em Proof of the Theorem \ref{trans}.}\\[-1.5em]

Equation (\ref{transcendental equation}) can be written as
\begin{equation}
\sigma e^{\sigma \mathrm{T}}\cdot e^{j\omega \mathrm{T}}+\omega e^{\sigma \mathrm{T}}\cdot e^{j(\omega \mathrm{T}+\frac{\pi}{2})}+\lambda=0,
\end{equation}
which is equivalent to
\begin{subequations}\label{transcendental equation2}
	\begin{align}
	\label{trans1}
	&\sigma e^{\sigma \mathrm{T}}\cos(\omega \mathrm{T})-\omega e^{\sigma \mathrm{T}}\sin(\omega \mathrm{T})=-\lambda,\\
	\label{trans2}
	&\sigma e^{\sigma \mathrm{T}}\sin(\omega \mathrm{T})+\omega e^{\sigma \mathrm{T}}\cos(\omega \mathrm{T})=0.
	\end{align}
\end{subequations}
It is possible to prove that if $(\sigma_i,\omega_i)$ is a pair of roots of equation (\ref{transcendental equation2}), 
then it has a pair of roots $(\sigma_i,-\omega_i)$.
Next, we first introduce how $\mathrm{T}^\star$ is derived, and then prove the validity of this theorem.

Let $\sigma=0$, then $s=j\omega$, and equation (\ref{transcendental equation}) is reduced to 
\begin{equation}
\omega \cdot e^{j(\omega \mathrm{T}+\frac{\pi}{2})}+\lambda=0.
\end{equation}
Without loss of generality, let $\omega\geq0$ and we can get
\begin{equation}
\left\{
\begin{split}
&\omega=\lambda\\
&\omega \mathrm{T}+\dfrac{\pi}{2} =\pi+2k\pi
\end{split}
\right.,
\end{equation}
so
\begin{equation}
\mathrm{T}=\dfrac{\pi/2+2k\pi}{\lambda}.
\end{equation}
Take $k=0$, then $\mathrm{T}^\star=\dfrac{\pi}{2\lambda}$.

In the following proofs, we will discuss the cases where $\omega\neq0$ and $\omega=0$, respectively.

When $\omega\neq0$, it can be inferred from equation (\ref{trans2}) that $\omega \mathrm{T}\neq k\pi$, $k=0,\pm 1,\pm 2,\cdots$. So equation (\ref{trans2}) can be expressed by 
\begin{equation}\label{exp_sigma}
\sigma=-\dfrac{\omega}{\tan(\omega \mathrm{T})}.
\end{equation}
Without loss of generality, let $\omega>0$ when $\omega\neq0$.

In the case of $\mathrm{T}<\mathrm{T}^\star$, assume that $(\sigma_i,\omega_i)$ is a pair of roots of equation (\ref{transcendental equation}), and $\sigma_i>0$. Based on equation (\ref{transcendental equation2}) we can get that
\begin{equation}\label{amplitude}
e^{\sigma \mathrm{T}}\sqrt{\sigma^2+\omega^2}=\lambda.
\end{equation}
It is clear that $\omega_i<\lambda$ when $\sigma_i>0$. Then
\begin{equation}
0<\omega_i \mathrm{T}<\lambda \mathrm{T}^\star=\lambda \dfrac{\pi}{2\lambda}=\dfrac{\pi}{2},
\end{equation}
so $\sigma_i<0$ according to equation (\ref{exp_sigma}), which contradicts the assumption above. To sum up, equation (\ref{transcendental equation}) has all roots in the open left half plane when $\mathrm{T}<\mathrm{T}^\star$.

Under the circumstances of $\mathrm{T}>\mathrm{T}^\star$, we want to show that equation (\ref{transcendental equation}) has at least one pair of roots $(\sigma_i,\omega_i)$ satisfying $\sigma_i>0$.
Substituting equation (\ref{exp_sigma}) into equation (\ref{trans1}), we can get the equivalent expression of equation (\ref{transcendental equation}),
\begin{equation}\label{exist}
\dfrac{x}{\sin x}=\lambda\mathrm{T}\cdot e^{\frac{x}{\tan x}}\\
\end{equation}
Let $\mathrm{f}(x)=\dfrac{x}{\sin x}-\lambda\mathrm{T}\cdot e^{\frac{x}{\tan x}}$, it is obvious that $\mathrm{f}(x)$ is continuous over $x\in(\dfrac{\pi}{2},\pi)$ and
\begin{equation}
\lim\limits_{x \to {\frac{\pi}{2}^+}} \mathrm{f}(x)=\dfrac{\pi}{2}-\lambda \mathrm{T}<\dfrac{\pi}{2}-\lambda \mathrm{T}^\star=\dfrac{\pi}{2}-\lambda \dfrac{\pi}{2\lambda}=0,
\end{equation}
\begin{equation}
\lim\limits_{x \to {\pi^-}} \mathrm{f}(x)=+\infty-\lambda \mathrm{T}\cdot 0^+ =+\infty-0>0.
\end{equation}
So there must exist a real number $x\in(\dfrac{\pi}{2},\pi)$ satisfying $\mathrm{f}(x)=0$ when $\mathrm{T}>\mathrm{T}^\star$. That is to say, if $\mathrm{T}>\mathrm{T}^\star$, then there exists a pair of roots $(\sigma_i,\omega_i)$ of equation (\ref{transcendental equation}), which satisfying $\omega_i\mathrm{T}\in(\dfrac{\pi}{2},\pi)$ and $\sigma_i>0$.

When $\omega=0,$ equation (\ref{transcendental equation}) is equivalent to
\begin{equation}\label{transcendental equation_special}
\sigma\cdot e^{\mathrm{T}\sigma}=-\lambda.
\end{equation}

If $(\sigma_i,0)$ is a pair of roots of equation (\ref{transcendental equation}), it is clear that $\sigma_i<0$ based on equation (\ref{trans1}).
Therefore, we only need to prove that equation (\ref{transcendental equation_special}) has no real roots when $\mathrm{T}>\mathrm{T}^\star$. Let $\mathrm{g}(\sigma)=\sigma\cdot e^{\mathrm{T}\sigma}$, it is easy to prove that $\mathrm{g}(\sigma)$ takes the minimum value when $\sigma=-\dfrac{1}{\mathrm{T}}$, and the minimum value is $\mathrm{g}_{\mathrm{min}}=\mathrm{g}(-\dfrac{1}{\mathrm{T}})=-\dfrac{1}{\mathrm{T}e}$. If $\mathrm{T}>\mathrm{T}^\star$, then
\begin{equation}
\mathrm{g}_{\mathrm{min}}=-\dfrac{1}{\mathrm{T}e}>-\dfrac{1}{\mathrm{T}^\star e}=-\dfrac{2}{\pi e}\lambda>-\lambda,
\end{equation}
so equation (\ref{transcendental equation_special}) has no real roots. This ends the proof.

\end{document}